# Big Data and Business Intelligence: Debunking the Myths

**Abstract**
Big data is one of the most discussed, and possibly least understood, terms in use in business today. Big data is said to offer not only unprecedented levels of business intelligence concerning the habits of consumers and rivals, but also to herald a revolution in the way in which business are organized and run. However, big data is not as straightforward as it might seem, particularly when it comes to the so-called dark data from social media. It is not simply the quantity of data that has changed; it is also the speed and the variety of formats with which it is delivered. This article sets out to look at big data and debunk some of the myths that surround it. It focuses on the role of data from social media in particular and highlights two common myths about big data. The first is that because a data set contains billions of items, traditional methodological issues no longer matter. The second is the belief that big data is both a complete and unbiased source of data upon which to base decisions.

## 1    Introduction

As the financial Times notes, big data, the unimaginably vast quantities of data that flow relentlessly from web sites, databases, information systems, mobile devices, social networks and sensors, is one of the most hyped, and one of the most confusing, business terms in use today (Laney, LeHong & Lapkin, 2013). It is said that big data will not only give businesses unprecedented insights into their customer's buying habits and their own internal processes, it is also claimed that it will herald a management revolution where technology replaces human judgement, enabling businesses to take better decisions, more quickly, and provide value for their customers in new and unimagined ways (McAfee & Brynjolfsson, 2012).

Dark data (Laney et al., 2013), that is data that can only be made visible through the processing of the data present in social media sites such as Facebook, YouTube, and Twitter, holds a special promise in this respect. Firms use social media to interact with their customers, to build their brand's identity, as well as to monitor their rivals. Social media sites can attract hundreds of millions of visitors and grow so quickly that statistics about their use becomes outdated before they reach the page. The growth of mobile communications, such as mobile telephones and tablets, combined with 'the internet of things' where internet-enabled devices exchange data without human intervention, has further contributed to the rate of growth of such data.

These large data sets seem to offer the prospect of access to forms of knowledge that were previously thought to be unobtainable; insights that were once thought of as difficult, now seem to be readily available. The value of such data to business appears to be unquestionable and few, if any, would argue that it should be ignored; the real problem appears to be how to make best use of it.

Businesses have always sought to glean intelligence from data and used it to gain competitive advantage; today business intelligence has developed into a wide range of activities that businesses undertake to understand their internal and external environment. However, as we shall see, the claims of unparalleled accuracy and objectivity made by the advocates of big data (Anderson, 2008) may not always be what they seem. Part of the problem lies not only with dealing with the volume of data, but also with the speed with which that data is produced and with variety of formats that are used to store and transmit it. Part of the problem lies with the lack of transparency behind the methods with which the data is collected and the complexity of the subsequent processing. Finally, for social media in particular, part of the problem lies with the source of the data: human beings.

In the sections that follow, we will first look at the impact of big data on business and at how business might use this data for business intelligence. We follow this with a more detailed examination of the phenomenon of big data, placing particular emphasis on the web as a source of data and the importance of the way in which that data is structured. We then turn to some of the ways in which businesses have tried to make effective use of this data, termed big data analytics. We examine some of the challenges posed by big data in general and the data produced by social media in particular. Finally, we conclude by examining some of the myths that surround big data and big data analytics.

## 2    Business Intelligence, Analytics and the Internet

### 2.1    Big Data and its Impact on Business

As we noted in the introduction, the impact of big data is undeniable; newspapers and academic journals are full of anecdotes and case studies that illustrate the value of such data to businesses. For example, McAfee and Brynjolfsson (2012) contrast a physical book shop, which can keep track of which books are sold and, if they have a loyalty program, can link some of those sales to individual customers, with an on-line store, such as Amazon. On-line stores, can not only track, with almost total accuracy, what was sold, to whom and when, they can also track what else they looked at; how they navigated their way through the web site and how they were influenced by promotions and special offers; furthermore, they can then use this data to predict what a customer might like to buy next.

However, some argue that the impact of big data goes beyond this. McAfee and Brynjolfsson (2012) also claim that big data heralds a revolution in the way that businesses are managed and managers are rewarded. They argue that when data is scarce, it makes sense for highly placed people to take decisions based on their intuition: the experience they have built up and the patterns they have internalized over their careers. Big data, they argue, will spell death for HiPPOs - the highest-paid person's opinions - as executive decisions become truly data driven. Some go further still and claim that big data will make whole swaths of human knowledge obsolete. For example, Anderson states "Out with every theory of human behavior ... Who knows why people do what they do? The point is they do it, and we can track and measure it with unprecedented fidelity" (Anderson, 2008)

### 2.2    Business Intelligence

The ability of a business to make use of the data that is available to it is sometimes termed business intelligence; the term was first popularized by Luhn (1958) who used it to describe the abstracting, encoding and archiving of internal documents and their dissemination using 'data-processing machines'. Later, the emphasis changed, and by the 1980s the ability to convert raw data into useful information for decision making was more highly stressed. Today the term business intelligence is used to cover a range of activities including competitor intelligence; customer intelligence; market intelligence; product intelligence; strategic intelligence; technological intelligence and even business counterintelligence. Consequently, the Gartner group now choose to describe business intelligence as simply "an umbrella term that includes the applications, infrastructure and tools, and best practices that enable access to and analysis of information to improve and optimize decisions and performance" (Gartner, 2013).

### 2.3    Business Intelligence and the Internet

The growth of the internet at the turn of the last century provided businesses with a wealth of new data that could be used for business intelligence. 'The Web' is said to be the largest publicly accessible data source in the world; Goggle's search engine alone has indexed more than 45 billion web sites (worldwidewebsize.com, 2015). Twitter posts in excess of 500 million tweets a day (internetlivestats.com, 2015); Facebook claims to have more than 936 million active users a day (Facebook.com, 2015) while YouTube claim to have more than 1 billion unique visitors each month and more than 6 billion hours of video: almost an hour of video for every person on Earth (YouTube.com, 2015).

Initially however, the internet was simply seen as a way to increase operational and financial efficiency by providing firms a new channel to deal with their customers and suppliers. Consequently, organizations began to invest in internet technologies simply as a means of communicating with suppliers and increasing their customer base. Nevertheless, as online markets began to grow, customers began to use the internet in new ways: to express their opinions, or to seek the opinions of others', about the products and services that were on offer. According to Nielsen (2012), of those customers who engage with companies through social media channels such as Facebook or YouTube, 70 per cent do so to hear others' experiences, 65 per cent do so to learn more about brands, products or services; while 50 per cent do so to express concerns or make complaints.

On-line reputation now makes a clear impact on the bottom line and a range of consumer review and comparison sites, such as TripAdvisor.com, have grown up to meet this demand. A study by The Kelsey Group and comScore (Kelsey, 2007), showed that consumers were willing to pay up to 20 per cent more for services rated by other customers as "5-star" in online reviews. Firms now actively encourage the users of social media to create reviews, initiate discussions, and make comments. A report by Burson-Marsteller Research on the Fortune global top 100 corporations' use of social media (Burson-Marsteller, 2012) showed that, in 2012, 87 used at least one social media platform, an increase of 8 percent from 2010. Recent research (Barnes, Lescault & Augusto, 2014) indicates that in 2014 93 per cent of the Fortune top 500 corporations now use social media tools.

**2.4   Business Analytics and Business Intelligence**
Clearly, business intelligence generated from big data could be of immense value; however, the current generation of analytics, the term coined for the analysis of web-based data in the early days of the internet, are unable to cope. To understand why this is the case, we need to examine the phenomenon of big data in more detail. To do this we will use the "Three Vs" definition - volume, velocity, and variety - first developed to describe changes related to the growth of e-commerce. Since then, others have attempted to add extra 'Vs" such as value, veracity and viability but, taken as a whole, these three dimensions are sufficient to provide a comprehensive picture of what big data is and to highlight its implications for the gathering of business intelligence.

# 3   Big Data
Most intuitive definitions of big data focus on the volume of data that is being produced, often measured in terms of tera ($10^{12}$), peta ($10^{15}$) or exa ($10^{18}$) bytes or, more colloquially, by making comparisons to a more tangible repository of data, such as. "X number of Libraries of Congress" (Johnston, 2012). Some claim we are entering The Petabyte Age (Anderson, 2008) while others prefer to talk of how many exabytes of data are produced each day (McAfee & Brynjolfsson, 2012). However, while volume is undoubtedly one aspect of big data, it is probably the least troublesome; as technology develops, what was big in the past will be normal tomorrow and probably thought of as quite small in the future. Consequently, to understand what makes big data different, we also need to consider the dimensions of velocity and variety; below we briefly review each in turn.

## 3.1 Volume

According to Hendler (2013), the term volume originated to describe the amount of data held in large organizational databases. As businesses go about their work, they inevitably generate data. As long ago as the 1980s Zuboff (1988) noted that as information systems automate organizational processes they also produce new information, making activities and events that were previously unseen, visible. For example, RFID data from supply chain applications has the potential to make each stage in a product's journey visible, no matter where the product is physically located. The volume of data that is generated within organizations will continue to grow inexorably as long as businesses use computers to manage their daily operations and engage in data gathering to support these activities.

More recently, however, the discussion has shifted from internal to external data, such as that found in web platforms. The volume of data available from the web has increased dramatically thanks to technologies like data streaming as well as everyday activities such as sending videos, pictures, or text messages. Recent developments, for example context aware applications that provide data about what users are doing, where they are located, who they are with and even, in the case of devices such as activity trackers, physiological data, have also contributed to this trend.

Much of this data is available to businesses through application programming interfaces (APIs); this, in conjunction with the high adoption rates, means that businesses are now able to access an enormous volume of data about their customers, potential new customers, the market, and their competitors.

## 3.2 Velocity

Whereas volume refers to what might be thought of as a 'stock' of data, velocity refers to rate at which that stock changes, for example, the speed at which data is generated, the frequency at which it is updated or the rate at which it is delivered. Examples of high velocity data include financial data from stock markets, real time data from sensors and video cameras, and clickstream data generated by visitors to online stores. In extreme cases, such as streamed data, both the generation and delivery of data is, effectively, continuous.

Those who are more agile and are the first to observe and exploit opportunities can gain significant competitive advantages (McAfee & Brynjolfsson, 2012); however, dealing with data velocity involves more than simply having sufficient bandwidth. An area of particular interest to business is being able to reduce the latency between the time that the data is created and the time that it is available to decision makers. In the case of internet users in particular, real time, or close to real time, data can provide knowledge about incipient market trends as well as highlighting emergent issues related to a brand's reputation.

Although a great deal of high speed data, such as Twitter's infamous 'firehose', is, in theory, available to business through streaming APIs, a major challenge is to decide what data should be saved, and what can be lost; at present, most businesses are only able to view this type of data through a brief (2 to 10 minute) sliding window (ScaleDB, 2015).

## 3.3 Variety

Although perhaps not as immediately obvious as volume or velocity, in many ways variety poses the biggest problem for the analysis of big data. Variety refers to the number of different sources that data can come from and the formats, structures, and semantics that are associated with them (Structure refers to both the format in which the data is stored, such as the number and length of fields, and, more crucially, the semantics that need to be associated with those fields. For a computer to be able to process data in a way that makes it valid and meaningful for human beings, the data first needs to be codified, that is a semantic value - effectively a meaning - has to be allocated to each item of data (Kimble, 2013).

Exhibit 1). The problem comes about because each different data source needs to be processed in a different way; therefore, although the data exists, it may not be structured in a way that makes it usable.

Structure refers to both the format in which the data is stored, such as the number and length of fields, and, more crucially, the semantics that need to be associated with those fields. For a computer to be able to process data in a way that makes it valid and meaningful for human beings, the data first needs to be codified, that is a semantic value - effectively a meaning - has to be allocated to each item of data (Kimble, 2013).

**Exhibit 1 Examples of Data Variety, adapted from Hurwitz, Nugent, Halper and Kaufman (2013)**

|  | Structured data | Unstructured data |
|---|---|---|
| Machine generated | Sensor data: Data from RFID tags, smart meters, medical devices, GPS data, or any sensors that automatically records data in a pre-defined way. | Satellite images: Including weather data or movement of tectonic plates, etc. |
| | Web log data: Operational data from servers, applications, network routers and so on, that collect data about their activity. | Photographs and video: Including security, surveillance, traffic video, etc. |
| | Financial data: Financial systems that generate data for stocks, bonds, and so on, on daily, hourly or real time basis. | Radar or sonar data: Including vehicular, meteorological, and oceanographic seismic profiles. |
| Human generated | Input data: Any kind of data that humans input into a computer. For example forms, CRM systems, survey and questionnaires, etc. | Internal textual data of an organization: e-mails, logs, survey results, reports, etc. |
| | Click-stream data: Data generated by human's interactions with web sites. | Social media data: Data from social media platforms such as Facebook, YouTube, Twitter, LinkedIn, or Flickr. |
| | Data related to virtual environments: Movement and actions of users in virtual worlds, such as SecondLife. | Mobile data: Including data such as videos, pictures, text messages, and location. |

In many ways, the current situation is similar to that faced in the early days of information systems when businesses needed to deal with data that had been generated by isolated pieces of software that had been built, in an uncoordinated way, to solve a variety of different problems. The solution then was the development of relational databases, where all of the different formats and semantics could be combined under one master data schema; however, as we shall see later, this will not provide the solution to all of the problems associated with big data.

Before turning to the issue of the analysis of big data, it should be noted that although the term 'unstructured', as in Structure refers to both the format in which the data is stored, such as the number and length of fields, and, more crucially, the semantics that need to be associated with those fields. For a computer to be able to process data in a way that makes it valid and meaningful for human beings, the data first needs to be codified, that is a semantic value - effectively a meaning - has to be allocated to each item of data (Kimble, 2013).

Exhibit 1, is often used to characterize data coming from certain sources. Strictly speaking, this is inaccurate. Data cannot be truly unstructured; some sort of structure must exist, either as a result the way it was produced, or the way it was consumed. However, something that is easy for a human to understand may pose severe difficulties for a machine; unstructured data is therefore a term that is usually used to describe data where the information content of that data is not readily amenable to automated analysis.

Having established the potential value of big data to business and having gained a better understanding of what big data is and what it consists of, we now return to the problem of extracting useful business intelligence from big data.

## 4   Big Data Analytics

Analytics and business intelligence are clearly related, however extracting business intelligence from big data is not as straightforward as it might seem. Below we review some of the issues associated with big data analytics. To do this we adopt the categorization of big data analytics produced by Chen, Chiang and Storey (2012) who refer to 3 different approaches to the analysis of big data: BI&A 1.0 (Business Intelligence and Analytics 1.0), BI&A 2.0 and BI&A 3.0. In addition, we will also draw a distinction between 'unstructured' data, principally data from social media, and other more structured forms of big data.

### 4.1   A Simple Typology of Business Intelligence and Analytics

BI&A 1.0 has its roots in relational databases, statistical techniques, and data mining techniques developed in the 1970s and 1980s. The data it deals with is mostly structured, internal data that has been collected by companies and stored in commercial, relational database systems. Chen et al. (2012) note that most of the data processing and analytical technologies for BI&A 1.0 have already been incorporated into the commercial business intelligence packages offered by major IT vendors.

BI&A 2.0 began to emerge at the turn of the century as businesses began to move on-line and interact with their customers directly. A vast amount of company, industry, product, and customer information can be gathered using various text and web mining techniques. In addition to information held in traditional databases, detailed, user-specific logs can be collected through cookies and server logs that can be used to guide web site design, and product placement (Ting, Clark & Kimble, 2009). Similarly, the analysis of customer transactions can be used to help understand market structure, and generate product recommendations. Although proprietary solutions exist, Chen et al. (2012) note that at present, apart from for basic query and search capabilities, no advanced analytics for unstructured data exist in commercially available business intelligence packages.

Finally, Chen et al. (2012) frame their discussion of BI&A 3.0 around the increasing use of mobile devices, such as the iPad, iPhone, and smart phones, and the development of ubiquitous computing, where devices such as televisions and motor cars contain embedded processors. Such mobile, Internet-enabled devices, they argue, will soon be used to support location-aware, person-centered, context-sensitive services. They also note that, at present, no commercial BI&A 3.0 systems currently exist and academic research is still in an embryonic state.

### 4.2   Analytics and More Structured Forms of Data

One of the major problems faced by BI&A 2.0 is dealing with the volume, velocity, and variety of big data. However, although technological solutions to these problems may be in sight, simply because we are able to process large amounts of data, this does not mean that that data will be either relevant, or useful.

Boyd and Crawford (2012) point out that Internet sources are prone to outages and losses, and that these gaps and errors tend to be magnified when several data sets are merged together. Corruption and loss of data are almost inevitable when dealing with large volumes of high velocity data; big data is not delivered into the hands of analysts pristine and ready for use, it first needs to be cleaned and conditioned to make it suitable for processing. Ekbia, Mattioli, Kouper, Arave, Ghazinejad, Bowman, Suri, Tsou, Weingart, and Sugimoto (2015) point out that this, combined with the opaque and under documented way in which data is gathered raises doubts about the supposed completeness and accuracy of big data.

From a slightly different viewpoint, Boyd and Crawford (2012) question the validity of the statistical techniques that are often used to analyze big data. To be able to use a statistical test to make claims about data, we need to know the properties of the data: where it came from, its distribution, and its weaknesses and biases. Simply because a data set contains billions of items does not mean that it is either random or representative. Without knowing how the data was collected and how it has been processed it is not possible to know if the assumptions upon which the tests are based have been violated. Ekbia et al. (2015) go further, claiming that because many of the tests that are used were designed to overcome the problems associated with small samples, their use with big data leads to apophenia: seeing patterns where none exist. They conclude that rather than removing the traditional dilemmas faced by analysts about what can legitimately be claimed from data, big data has actually made them worse.

If big data has exacerbated the problems of the analysis of more easily quantified, structured data, what does this mean for the analysis of the valuable, but less structured forms of data found in social media?

### 4.3 Analytics and Less Structured Data from Social Media

Despite the recent emergence of the notion of the internet of things, the content of the internet is still primarily created by people. Tim Berners-Lee, often credited the invention of the World Wide Web, said of his creation, "The Web is more a social creation than a technical one. I designed it for a social effect - to help people work together - not as a technical toy" (Berners-Lee & Fischetti, 1999, p. 123). The early web was characterized by static web pages providing one-way communication, often termed 'Web 1.0'. The technologies that dominate internet today, known as 'Web 2.0', allow the creation and modification of content by groups of people as well as the combination and reuse of data from different applications; the most prominent of these are social networks sites created to serve groups of people who share common interests.

Web 2.0 has changed the way people interact on-line and has led to the formation of what have become known as virtual communities. The growth in the use of social media is an almost inevitable consequence of this. Effectively, the web has become a medium for human communication, with all of the subjectivity, confusion, misunderstandings, misinterpretations, and deliberate deception this entails. In the context of our discussion, these communities form to share knowledge, opinions, and experiences about products and services. However, despite of the potential value of this information to business, according to Patterson (2012), existing analytics tend to be limited to quantitative assessments, such as how many times a brand is mentioned (Exhibit 2).

**Exhibit 2. Some Common Social Media Metrics**

| Metric | Description |
|---|---|
| Channel distribution | Calculated across several platforms to see which brands are the subjects of discussion, and on what platforms. |

| Metric | Description |
|---|---|
| Engagement | Indicates the level of involvement of users in the brand, usually measured by the number of likes, followers, shares, tweets, etc. |
| Geography | Indicates the geographical origin of comments based on information provided by users, or via IP addresses/GPS sensors. |
| Influencer ranking | A measurement of the popularity of users that create content referring to a specified brand; calculated on the number of connections that user has. |
| Sentiment | Indicates the attitude towards the brand using linguistic algorithms that identify positive and negative words. |
| Topic and theme detection | Information relating to a specific brand concerning the nature of a topic that was discussed; allows popular topics to be identified. |
| Volume of Posts | Indicates the number of items of user-generated content (e.g. blog posts, articles, or videos) that contain the name of the brand. |

Metrics based on simple counts of activities however are unlikely to provide any deeper understanding of the interactions that take place. Such measures treat social interactions as unproblematical quantitative data and risk oversimplifying the rich and dynamic nature of the communication that takes place. For example, Ekbia et al. (2015) cite the example of the BBC (British Broadcasting Corporation) who unveiled an initiative to "map the mood of the nation" by classifying twitter feeds according to eight basic human emotions. They ask,

> "Even if we assume that human emotions can be meaningfully reduced to eight basic categories (what of complex emotions such as grief, annoyance, contentment, etc.?) ... how does one differentiate the "happiness" of the fans of Manchester United after a winning game from the expression of the "same" emotion by the admirers of the Royal family on the occasion of the birth of the heir to the throne?" (Ekbia et al., 2015, p. 8)

In addition, these measures are blind to the playful, creative, unusual, and sometimes eccentric ways in which people use social media. The content of social media should give businesses access to information about their customer's opinions, ideas, thoughts, and feelings; however, moving beyond the generation of simple quantitative measures poses some difficult practical and philosophical questions.

### 4.4 Social Media and the Nature of Human Communication

The Austrian mathematician and philosopher Wittgenstein spent the first part of his life searching for stable, ideal meanings for words in an attempt to define the principles of language using logic. He later rejected the notion of a language in which words had meanings that were unique, identifiable, and stable, and instead claimed that language could only be understood in the context in which it was used. He argued that linguistic terms arise from social conventions created by people rather than by reference to some objective external reality. He saw language as a game where the rules were created as it was played; consequently, the only way to understand the rules, was to participate in the game. Marshall and Brady (2001) summarize this argument as follows, "Linguistic meaning is never complete and final … It is unstable and open to potentially infinite interpretation and reinterpretation in an unending play of substitution" (Marshall & Brady, 2001, p. 101).

Viewed in this way, the players of Wittgenstein's language game can seen as the communities whose practices provide the only fixed point against which the meanings given to words can be anchored. It is the observation that the semantics of words and images are inextricably rooted in life experience of the people who use them that poses the greatest problem for the analysis of data from social media.

Different communities will view the same thing in different ways and use different words or images to describe it. Similarly, the same words or images may have quite different meanings in different communities, for example, the meaning attached to the logos of established brands can be parodied to give it a quite different meaning. As Petty (2012) notes, while regular searches for the use of a brand name will uncover uses that spell the name correctly, they will not uncover misspellings that seek to parody or create a negative image for the brand.

**4.5    Analyzing Social Media**
As Chen et al. (2012) observed, research on BI&A 3.0 is still in an embryonic stage; this is particularly the case for social media. In the case of parodies of established brands, it is possible to use targeted image recognition and text analysis software to highlight the misuse of brand names and registered logos. However, the real value of the data from social media lies not in protecting established trademarks but in discovering new ideas and identifying emerging trends. However, if these trends and ideas are truly novel, and they are expressed using a new or unknown terminology, how could they even be identified using conventional approaches?

Currently there are a plethora of methods used for the analysis of social media including social network analysis, text and web mining, natural language processing and sentiment analysis, however the results of such analyses are often limited to the constraints of the particular analytical tool that was used or to a particular source of data (Milolidakis, Akoumianakis, Kimble & Karadimitriou, 2014b). Thus, for example, although social media users rarely restrict their activities to one platform, an analysis of Facebook using social network analysis may not recognize that the same user is cross posting similar material to Twitter. What is needed is not so much advances in technology but some form of methodological protocol that will allow us to combine data from a range of analytical tools and data sources so that changing patterns of meanings can be tracked across time and across media (Milolidakis, Akoumianakis & Kimble, 2014a).

Jones (2003), offers one such approach, based on traditional archeology, which he terms cyber-archeology. Jones set out to develop a methodology to study online public interactions that was not culture or time specific. He notes that the traditional archaeology provides us with a perspective that allows us to study the process of cultural change and to identify the slow impact of changes in a community's behavior over time (Jones & Rafaeli, 2000). He argues that cyber-archeology has a similar potential and claims that, like the excavation of archaeological Tells - the mounds of debris that accumulate around human settlements - the excavation of virtual Tells, the digital traces left by virtual communities, can tell us about what has taken place in those communities.

This approach has been adopted in a number of studies. For example, Milolidakis used this approach in studies relating to support groups for people with various types of cancer (Akoumianakis, Karadimitriou, Vlachakis, Milolidakis & Bessis, 2012) and also for the fan pages of Greek telecommunication companies on Facebook (Milolidakis et al., 2014a). He has also used this approach, again with telecommunication companies, in cross platform studies using Facebook and YouTube (Milolidakis et al., 2014b). It is important to stress however that, like traditional archaeology, this approach tends to be slow and labor intensive as much of the work involves the interpretation, rather than the automated processing, of data. Nevertheless, as Boyd and Crawford (2012) point out, as soon as an analyst starts to ask what the data means - regardless of the source - the process of interpretation by human beings begins.

Cyber-archeology is not a panacea; in the same way that modern archaeological techniques developed from the work of a few individuals in the 18th and 19th century, so more work is needed to develop protocols for the excavation of data from virtual settlements in the 21st. In addition, technological factors also limit the usefulness of this approach. For example, the weak and incomplete archiving associated with some social media and the limitations of the associated APIs limit the number of layers of context and meaning that can be uncovered. Thus, while Wikipedia maintains meticulous records of changes to its content, it only offers a basic APIs for exporting that content; Twitter on the other hand offers a range of ways to access content, but has only weak archiving facilities.

## 5   De-Bunking Big Data

We have seen that, on the one hand, big data appears to herald a revolution for businesses where it becomes possible to have unparalleled insights into customer's needs and the activities of competitors; it also seems set to herald a revolution in the ways in which businesses are run, with hard data rather than intuition driving decisions. Big data, analytics, business intelligence, and the Internet have aligned to usher in a brave new world.

On the other hand, we have seen that dealing with big data is not as straightforward as it might seem; it is not simply the quantity of data that has changed, it is also the speed and the variety of formats with which it is delivered. We have also seen that the analysis of data from social media, sometimes called dark data because the patterns it reveals are invisible to the human eye, poses particular problems because of the interpretive flexibility of words and images and the mischievous tendency of human language to morph and change over time.

There is no doubt that advances in technology will help to overcome some of these problems, particularly those associated with the handling of large volumes of high velocity data; it is also probable that some the problems associated with the range of formats in which data is supplied will be overcome in due course. However, like human language, standards and formats change and evolve over time and so called "standards wars" are a natural feature of competition as companies struggle to establish the preeminence of one standard over another to gain or maintain a dominant position in the market.

What then does this have to say about the use and value of big data to business? Having reviewed the limits of big data, there is clearly a need to de-bunk some of the myths that surround it.

Firstly, big data does not provide easy answers. Boyd and Crawford (2012) comment that Anderson's sweeping dismissal of all other theories reveals an undercurrent that is present in many discussions of big data where all other forms of analysis are scorned. From a slightly different perspective, Spender (2014), commenting on the role of managerial judgment, observes that in the past managers were seen as people who had to manage under determined situations. Now, big data and the trend towards IT-intensive practices means that managers are seen as people who have to deal rationally with determinable situations; an approach that is only viable, he claims, if we believe that what was under-determined in the past has now become fully determined, calculable, and forecastable.

Ekbia et al. (2015) argue that big data has led to a shift away from causal explanations towards predictive modeling and simulation; echoing the words of microbiologist Carl Woese they warn that while this might show us how to get there, it won't tell us where 'there' is. To those, such as Anderson (2008), who argue "Who knows why people do what they do? The point is they do it" this may seem to be of little importance. However, as we have seen, data taken out of context loses its meaning and value, and when large data sets are turned into mathematical models, data is inevitably decontextualized and reduced to what will fit into such models. The risk is that big data will provide accurate, but essentially meaningless, answers.

Secondly, there is a need to de-bunk the belief in the supposed objectivity of big data. We have already seen some of the technological and methodological reasons why big data may not be as complete and objective as it seems; something that should be apparent to even a casual user of social media. For example, the use of Facebook's like button, which is taken as an indication of approval, can easily be manipulated by offers such as "like our product and enter a draw to win a luxury holiday". Similarly Boyd and Crawford (2012) note that "Twitter does not represent 'all people', and it is an error to assume 'people' and 'Twitter users' are synonymous" (Boyd & Crawford, 2012, p. 669): some people have multiple accounts, while some accounts are used by multiple people; some people are not people at all but automated "bots" that generate content without the direct human intervention. Unless we accept the myth that big data sweeps away the need for methodology, observations such as these should give us cause to question the objectivity of the data we receive.

Similarly, one of the arguments for using big data from social media is that the data has all been made publicly available, so it is free and there is no need to ask permission to use it. Leaving to one side the issues of anonymity and who actually has access to the data, this issue still raises some potential legal and ethical concerns. Laney (2012) calls Facebook's users "the largest unpaid workforce in history" indicating that the average value of the data posted on Facebook is $ 81 per person. It is reasonable to assume that many of Facebook's users are not aware of the uses that will be made of what they have posted, nor of the profits, and other gains that will flow from it. The data contained in Facebook was created in a particular context; it is entirely possible that some users would not give their permission for their data to be used in a different context. Although they do not have the high profile of other issues currently, there is no doubt that legal issues concerning the use of data, as well as commercial judgements, will affect what is made available and to whom in the future, further undermining the supposed completeness and objectivity of big data.

To conclude, there is no doubt that, thanks to big data, we can now make predictions that are faster and more accurate than before and take, at least potentially, better informed decisions based on that data. However, it is equally clear that the blind enthusiasm with which some have taken up the cause of big data risks undermining these gains. Neither big data, nor technological wizardry alone, will provide all the solutions; rather we need to find ways to bring both human wisdom and technological prowess to bear on the complex of questions that surround big data.